\documentclass[twocolumn,prl,english]{revtex4-2}
\usepackage[T1]{fontenc,xcolor}
\usepackage[latin9]{inputenc}
\usepackage{verbatim}
\usepackage{amsmath,bm}
\usepackage{amssymb}
\usepackage{graphicx}
\usepackage{babel}
\usepackage{subfigure}

\makeatletter

\newcommand{\e}{\mathrm{e}}
\newcommand{\vecr}{\mathbf{r}}
\newcommand{\vecl}{{\bm \ell}}
\newcommand{\vecp}{\mathbf{p}}

\newcommand{\vecv}{\mathbf{v}}

\newcommand{\vecg}{\mathbf{g}}

\newcommand{\D}{\mathrm{d}}
\newcommand{\half}{\frac{1}{2}}

\makeatother

\begin{document}

\title{Chiral active fluids: what can we learn from the total momentum?}

\author{Tomer Markovich$^{1,2}$}
\email{tmarkovich@tauex.tau.ac.il} 
\author{Tom C. Lubensky$^{3}$}
\affiliation{
$^{1}$School of Mechanical Engineering, Tel Aviv University, Tel Aviv 69978, Israel \\
$^{2}$Center for Physics and Chemistry of Living Systems, Tel Aviv University, Tel Aviv 69978, Israel \\
$^{3}$Department of Physics and Astronomy, University of Pennsylvania, Philadelphia, Pennsylvania 19104, USA
}

\date{\today}

\begin{abstract}
Chiral active materials are those that break both time-reversal  symmetry and parity microscopically, which results in average rotation of the material's complex molecules around their center-of-mass (CM). These materials are far from equilibrium due to their local non-vanishing spin angular momentum. In this paper we show that, unlike passive fluids, the non vanishing spin angular momentum brings about a difference between the CM  momentum and the total momentum, which accounts for the momentum of all atoms that compose the complex rotating molecules. This is in stark contrast to equilibrium fluids where the CM stress and the total momentum are essentially equivalent. In fact, we find that generally the CM dynamics are insufficient to describe the dynamics of a chiral active material.
The total momentum, other than being experimentally accessible in simple rheological experiments, also imposes another constraint -- its stress must be allowed to be written in a symmetric way. We find that the latter imposes a relation between possible central-force interactions and spin-spin interactions, and constraints the ammount of {\it odd} viscosities in the system to the well-known odd (Hall) viscosity and the odd pressure. 
\end{abstract}
\maketitle


\section{Introduction}

Active materials~\cite{Marchetti2013}, in which energy is injected in the microscale to create directed mechanical motion, break time reversal symmetry (TRS) locally, and constantly dissipate heat to maintain such motion~\cite{PRX2021,fodor2022}. Much progress has been made in studying achiral active matter where usually force dipoles generate the directed motion~\cite{Ramaswamy2004,Joanny2015,Tjhung2013,Broedersz2011}. Exciting novel behavior  that cannot be seen in passive fluids has been observed, from motility-induced phase separation~\cite{MIPS} to negative apparent viscosity~\cite{PRL2019}, new collective behavior~\cite{Maitra2021,Chen2020} and pattern formation~\cite{Solon_pattern_2015}. This had implication on the development on new metamaterials~\cite{Bartolo2013,Bartolo2015,Spec2013} and on the understanding of biological systems~\cite{Sumi2024,Ramaswamy2005,Marchetti2013,Joanny2015,Broedersz2011}.


Recently, a new class of active materials attracted much interest: Chiral active materials, in which TRS and parity are broken locally~\cite{odd_review,Mandadapu2020,Rudi2022,MarLub2021,MarLub2024}. Quite generically this is a result of local injection of torques that also result in injection of angular momentum~\cite{NJP,furthauer2012}. Biology offers a plethora of examples of chiral active materials~\cite{Naganathan2014,Goldstein2009,Tabe2003,Petroff2015}, and recently these were also realized in synthetic materials~\cite{Snezhko2016,Maggi2015,Irvine2019,Tsai2005,Bartolo2016,odd_review}. 
A striking phenomenon in these chiral active materials is the appearance of odd viscosity~\cite{MarLub2021,MarLub2024,odd_review,Banerjee2017,Ganeshan2017}, which is a non-dissipative viscosity~\cite{odd_review,MarLub2021,MarLub2024,Abanov2021} that couples the two directions of shear~\cite{VitelliKubo}. Avron showed that, as Onsager predicted~\cite{Onsager,Mazur}, odd viscosity can appear when TRS is broken~\cite{Avron1998}, but it is only compatible with isotropy in 2D. Although Onsager predicted that breaking of TRS is sufficient in order for odd viscosity to appear, in all work we are aware of parity is also broken, whether in gases~\cite{Kagan-Mak}, plasmas~\cite{Braginskii,LLfluids}, quantum-Hall fluids~\cite{Hoyos2014,Read2009,Read2011,Bradlyn2012,Gromov2017}, or in chiral active materials due to local torques injected at the particle level~\cite{NJP,MarLub2021,VitelliKubo,Banerjee2017,Rudi2022,Andelman2021,Mandadapu2020}. Importantly, in structurally isotropic chiral active matter, odd viscosity also appears in 3D~\cite{MarLub2021,MarLub2024,Khain2022} where angular momentum breaks rotational symmetry~\footnote{More odd terms are allowed by symmetry in 3D materials with cylindrical symmetry~\cite{Khain2022}.}. 
%
Recent work on active materials propose new `odd' terms (that break TRS and parity)  that are allowed by symmetry, even in isotropic materials. In 2D these terms are the odd pressure that couples vorticity with pressure and odd torque that couples dilation with rotation~\cite{Abanov2021,VitelliKubo,odd_review}. 

When dealing with complex fluids in which the constituents are not point-like atoms but rather complex molecules composed of many atoms, the momentum of each molecule is not simply its center-of-mass (CM) momentum. For example, a rigid body has three transnational and three rotational degrees-of-freedom. In an isotropic complex fluid composed of rigid complex particles, the extra degrees-of-freedom are captured via the spin angular momentum (SAM) density in addition to the linear momentum. Then, the {\it total hydrodynamic momentum} $\vecg = \vecg^c + \half\nabla\times\vecl$ (hereafter referred to as {\it total momentum} for brevity), which also includes  rotational motion around the CM of the fluid molecules, captures the momentum of all atoms~\cite{MarLub2024} in the system in the hydrodynamic limit. The concept of total  momentum is not new. In the context of classical fields it was first introduced by Martin {\it et al.}~\cite{Martin} and it is also known in the quantum physics literature as the Belinfante-Rosenfeld relation~\cite{Bliokh2021}.

In fluid mechanics it is common to describe the velocity field dynamics using the Navier-Stokes equation, which combines the conservation of linear momentum and a constitutive relation between stress and gradients of velocity. Since SAM density (which is always present in non point-like molecules) generically relaxes fast compared to the velocity field (it has a finite relaxation time, i.e., it is non-hydrodynamic), conservation of total angular momentum requires the stress tensor to be symmetric in passive fluids. 
This is not the case in chiral active matter. In these materials, energy is injected locally via torques, which also results in injection of angular momentum at the microscopic scale. Then, the SAM relaxes to a non-vanishing value at hydrodynamic times and there is no guarantee that the CM stress is symmetric. Nevertheless, the total momentum stress tensor can always be written in a symmetric way~\cite{Martin,MarLub2021,MarLub2024,LLelasticity}.

This poses a conundrum: What is the fluid stress? Is it the CM stress or the total momentum stress? Or perhaps they are equivalent?  
Clearly, the total momentum stress, which accounts for the momentum of {\it all} atoms (in the hydrodynamic limit) is the actual force on the system boundaries, and is thus the stress that is accessible experimentally in a rheological experiment~\cite{MarLub2024}. The CM stress can be accessed using other microscopic measurements (e.g., mean-square-displacement of a tracer particle).
Martin et al.~\cite{Martin} argued that in passive fluids the difference between the total momentum and the CM stress tensors is a microscopic quantity that has no hydrodynamic effects. In Ref.~\cite{MarLub2024} we showed that this is not the case in chiral active materials using a microscopic  (although very general) model.

In this paper we derive the most general viscosity tensor phenomenologically using symmetry constraints and physical constraints arising from conservation laws. Specifically we utilize the notion of total momentum, for which a symmetric stress tensor can always be found~\cite{LLelasticity,Martin,MarLub2024}. This leads to constraints on the `odd' part of the viscosity tensor, thus relating seemingly separate terms, which constrain the type of interactions possible between particles in an odd fluid; two-body interaction between particles CM and spin-spin interactions are strongly coupled. 
As a consequence, at hydrodynamic times one can only expect to measure two odd coefficients: The famous odd or Hall viscosity and an odd pressure~\cite{Abanov2021}. At short times, before the  SAM relaxes, one might get access to the so-called odd torque~\cite{Abanov2021}, but stress measurements, that are always related to the flux of total momentum (rather than the stress associated with the CM momentum), only provide access to the odd viscosity and odd pressure (at all times).
%

As an answer to the conundrum posed above, we find not only that the two stresses (the CM stress and the total momentum stress) are not equivalent, even after relaxation of SAM, but also that the CM dynamics cannot fully describe the system (in general) as it does not conserve total angular momentum.

In the next section (Sec.~\ref{sec:model}) we define the model system and show the result obtained microscopically for a non-interacting fluid, Sec.~\ref{sec:pb}. In Sec.~\ref{sec:general_stress} we discuss the most general odd viscosity possible, including constraints from the structure of the total momentum, Sec.~\ref{sec:symm}. Then, we eliminate the angular momentum in Sec.~\ref{sec:elim} showing that the CM description is insufficient for chiral active fluids. We discuss our findings in Sec.~\ref{sec:discussion}

\begin{figure}
	\begin{centering}
		\includegraphics[scale=0.4]{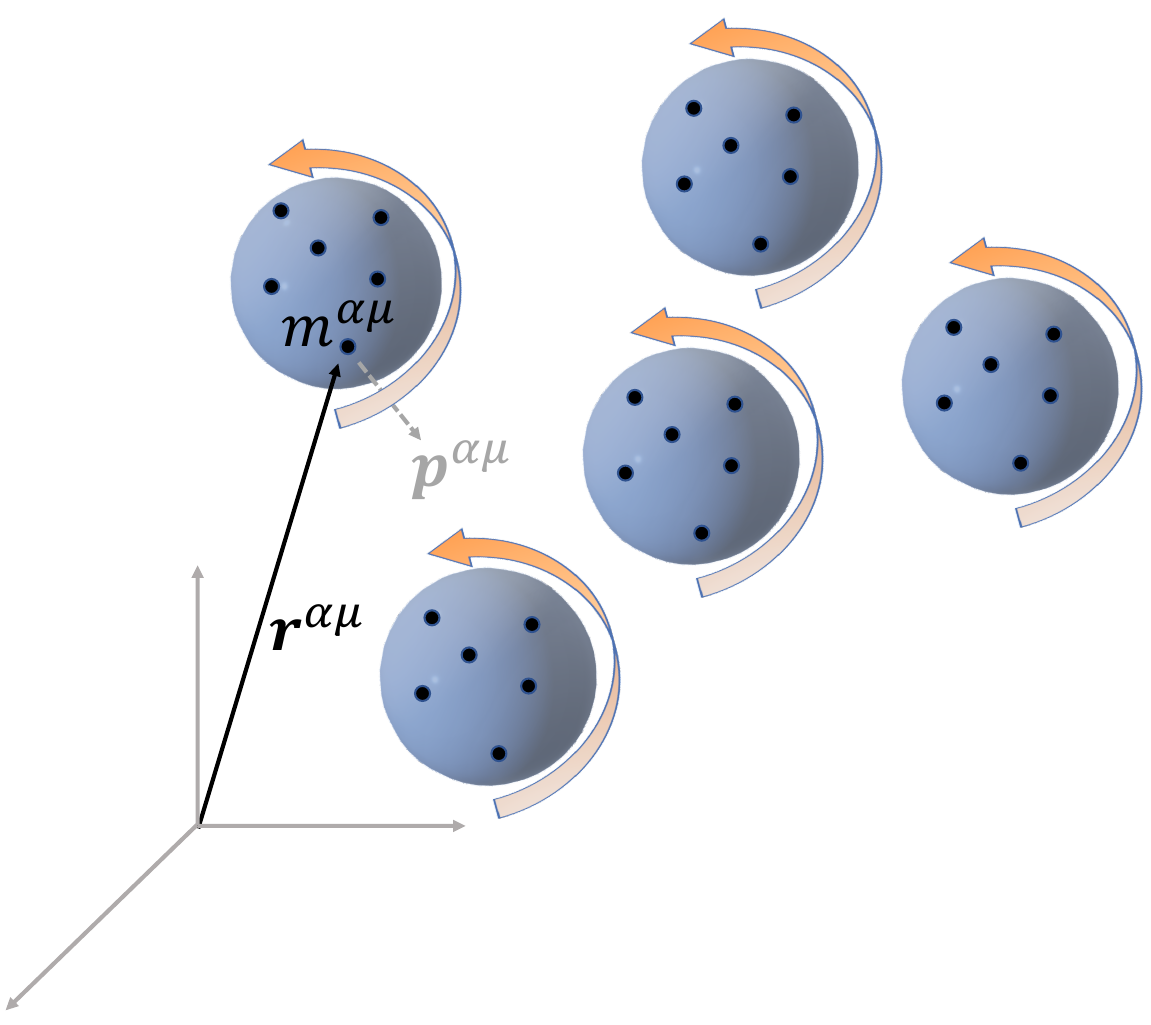}
		\par\end{centering}
	\caption{Schematic of our model system. Each molecule with CM at $\vecr^\alpha$ (large blue spheres) 
		is composed of some point-like particles (small black spheres) located at $\vecr^{\alpha\mu}$ with mass $m^{\alpha\mu}$ and momentum $\vecp^{\alpha\mu}$. In this work we focus on the 2D case in which the spin angular momentum direction is perpendicular to the surface (out of the page in the figure).
	}
	\label{fig:schematic} 
\end{figure}

\section{Model}\label{sec:model}

Consider an isotropic fluid composed of complex particles labeled by $\alpha$  (i.e., those that has at least two atoms labelled by $\mu$) as depicted in Fig.~\ref{fig:schematic}. The hydrodynamics of such isotropic fluid can be described using conservation of mass and momentum:
\begin{eqnarray}
\label{eq:continuity}
& &\dot\rho +\nabla_i \left(\rho v_i \right) = 0 \, , \\
\label{eq:NSE}
& &\dot{g}_i + \nabla_j(v_jg_j) = \nabla_j\sigma_{ij} + f_i \, ,
\end{eqnarray}
where $\hat\rho(\vecr,t) = \sum_{\alpha\mu} m^{\alpha\mu} \delta(\vecr-\vecr^{\alpha\mu})$ is the mass density and 
$\hat\vecg(\vecr,t) = \sum_{\alpha\mu} \vecp^{\alpha\mu} \delta(\vecr-\vecr^{\alpha\mu})$ is the total momentum density. Here $\rho(\vecr,t)$ and $\vecg(\vecr,t)=\rho(\vecr,t)\vecv(\vecr,t)$ are the coarse-grained fields, with $\vecv(\vecr,t)$ being the velocity, and $\boldsymbol{f}$ is an external force. Throughout this paper all forces that conserve total linear momentum (active or passive) are included in ${\bm \sigma}$, such that only $\boldsymbol{f}$ is allowed to break conservation of total linear momentum density ($\boldsymbol{f}$ can also break conservation of total angular momentum density as will be shown below). With this distinction, the stress associated with the total momentum, ${\bm \sigma}$, can always be written as a symmetric tensor up to a term $\half\varepsilon_{ij} \tau^{\rm ex}$ where $\tau^{\rm ex}$ is an external torque that breaks conservation of total angular momentum density~\footnote{The total momentum stress tensor can have antisymmetric terms that can be written as a divergence of a third-rank tensor, but these terms can be symmetrized~\cite{NJP,LLelasticity,furthauer2012}.}.

Most of the discussion in this article is relevant to odd fluids in any dimension. To make the discussion simpler and to highlight the essential physics, we focus on 2D fluids. As was discussed in Refs.~\cite{MarLub2021,MarLub2024}, the total momentum in an isotropic fluid can be written as
\begin{eqnarray}
\label{eq:total_momentum}
g_i  = g^c_i + \frac{1}{2} \varepsilon_{ij} \nabla_j \ell  \, , 
\end{eqnarray}
where $\vecg^c(\vecr,t)=\rho(\vecr,t)\vecv^c(\vecr,t)$ is the coarse-grained CM momentum and $\hat\vecg^c(\vecr,t) = \sum_{\alpha} \vecp^{\alpha} \delta(\vecr-\vecr^{\alpha})$. Here,  $\vecr^\alpha$ is the position and $\vecp^{\alpha}  = \sum_\mu \vecp^{\alpha\mu} \delta(\vecr-\vecr^{\alpha})$ the momentum of the molecule $\alpha$ center of mass.  $\ell=\rho\tilde{I}\Omega$ is the SAM density where $\tilde I$ is the moment of inertia per unit mass (the moment of inertia density is $I=\rho\tilde{I}$), and $\Omega$ is the angular velocity. 

The CM dynamics obeys Eqs.~\eqref{eq:continuity}-\eqref{eq:NSE} with ${\bm \sigma} \to {\bm \sigma}^c$ and $\vecv \to \vecv^c$, but because the CM stress tensor is not the total momentum flux it can have antisymmetric parts as will be discussed below. 

\subsection{Poisson-bracket results for non-interacting spinning particles} \label{sec:pb}

When considering a system of non-interacting spinning particles, it has been shown in Ref.~\cite{MarLub2024} that using the Hamiltonian $H = \int\D\vecr\left[  \left(\vecg^c\right)^2/(2\rho) + \vecl^2/(2I) \right]$ and the Poisson-bracket formalism~\cite{Lubensky2003,Lubensky2005,LubenskyBook}, odd viscosity naturally emerges in the total momentum stress tensor:
\begin{eqnarray}
\label{eq:pb_total_stress}
\sigma_{ij} = -P\delta_{ij} + \left( \eta^{e}_{ijkl}  + \eta^{o}_{ijkl} \right)\nabla_l v_k  \,  ,
\end{eqnarray}
where $P(\rho)$ is the pressure and
\begin{eqnarray}
\label{eq:even_viscosity}
& &\eta^e_{ijkl} = \lambda \delta_{ij}\delta_{kl} + \mu \left( \delta_{il}\delta_{jk} + \delta_{ik}\delta_{jl}\right)\, ,\\
\label{eq:odd_viscosity}
& &\eta^o_{ijkl} = -\frac{\ell}{4} \left( \gamma^o_{ijkl} - 2\varepsilon_{lk}\delta_{ij}  \right) \, ,
\end{eqnarray}
with the usual~\cite{Avron1998,MarLub2021,LLfluids,Banerjee2017} odd viscosity tensor 
\begin{eqnarray}
	\label{eq:odd_tensor}
	\gamma^o_{ijkl} = \varepsilon_{jl} \delta_{ik}  +  \varepsilon_{ik} \delta_{jl} + \varepsilon_{il} \delta_{jk}  +  \varepsilon_{jk} \delta_{il} \, ,
\end{eqnarray}
such that the odd viscosity coefficient is $-\ell/4$, and $\lambda$ and $\mu$ are constants.

On the same footing, it has been found~\cite{MarLub2024} that the CM stress tensor is quite different,
\begin{eqnarray}
\label{eq:pb_cm_stress}
\sigma^c_{ij} &=& -P\delta_{ij} + \eta^e_{ijkl} \nabla_l v^c_k  +  \frac{\Gamma}{2} \varepsilon_{ij} \left(\Omega - \omega^c\right) \, ,
\end{eqnarray}
where $\omega^c = \half\nabla\times\vecv^c$ is the rotation vector.
The last term in this equation is a dissipative antisymetric stress that tends to make the fluid rotate in the same angular velocity as its constituents.


\section{General ``odd'' stress}\label{sec:general_stress}

Up to this point all quantities were defined microscopically. We proceed with a phenomenological approach~\cite{Abanov2021,VitelliKubo, NJP} to write the most general CM stress tensor. To first order in gradients it can be written as
\begin{eqnarray}
\label{eq:general_cm_stress}
\nonumber\sigma_{ij}^c &=& -P(\rho,\ell^2,(\vecv^c)^2)\delta_{ij} + \eta_{ijkl}(\rho,\ell,(\vecv^c)^2) \nabla_l v^c_k \\
&+& v^c_k\nabla_l \chi_{ijkl}(\rho,\ell,(\vecv^c)^2) \, .
\end{eqnarray}
The first term in the right-hand-side is the pressure, which is usually only a function of the density, but it could in principle include other terms that depend on $\ell^2$ or $\vecv^2$. The second term is the usual viscosity tensor, where usually only linear terms in $|\vecv|$ are kept. The third term breaks Galilean invariance and is usually absent in hydrodynamic formulations. Here we include this term explicitly because we have shown in Ref.~\cite{MarLub2021} that while the total momentum dynamics obey Galilean invariance (up to terms arrising from $\boldsymbol{f}$), there is no guarantee that the CM dynamics obeys it. Hereafter we only keep terms to linear order in $|\vecv|$. A fourth rank tensor in an isotropic 2D system has up to six independent coefficients~\cite{VitelliKubo,Abanov2021}:
\begin{eqnarray}
\label{eq:general_viscosity}
\nonumber& &\eta_{ijkl} = \eta^e_{ijkl} + \frac{\eta^o}{4} \gamma^o_{ijkl} + \frac{\Gamma}{4} \varepsilon_{ij}\varepsilon_{kl} \\
& &\qquad\,\,-\, \eta^A\varepsilon_{ij}\delta_{kl} - \eta^B \varepsilon_{kl}\delta_{ij} \, ,
\end{eqnarray}
and similarly for $\chi_{ijkl}$. The terms $\eta^B$ and $\eta^A$ are sometimes refer to as odd pressure and odd torque, respectively~\cite{Abanov2021}. The $\Gamma$ term is symmetric under $ij \leftrightarrow kl$ and is therefore dissipative (as is the regular dissipative viscosity $\eta^e_{ijkl}$ that has two coefficients, see Eq.~\eqref{eq:even_viscosity}). This term is the one that tends to make the fluid rotate with the same angular velocity as its constituents, and thus appear always in the form $\sim (\Omega-\omega^c)$~\cite{Tsai2005,Banerjee2017,MarLub2021,MarLub2024,VitelliKubo}. 
To conclude, the CM stress tensor reads
\begin{eqnarray}
\label{eq:cm_stress}
& &\sigma_{ij}^c = -P\delta_{ij} + \varepsilon_{ij}\frac{\Gamma}{2} (\Omega-\omega^c) + v^c_k\nabla_l \chi_{ijkl}\\
\nonumber& &+ \left[ \eta^e_{ijkl} +  \frac{\eta^o}{4} \gamma^o_{ijkl} - \eta^A\varepsilon_{ij}\delta_{kl} - \eta^B \varepsilon_{kl}\delta_{ij} \right] \nabla_l v^c_k  \, .
\end{eqnarray}

The dynamics of the SAM density can be generally written as~\cite{NJP}
\begin{eqnarray}
\label{eq:tam}
\dot{\ell} + \nabla_j (v^c_j \ell) = \tau^{\rm ex} - \varepsilon_{ij}\sigma^c_{ij} + \nabla \cdot \boldsymbol{C} \, ,
\end{eqnarray}
where $\boldsymbol{C}$ is a couple-stress (the analogue of stress for angular  momentum) and it may contain active and/or passive terms~\cite{NJP}. Similar to $\boldsymbol{f}$, here only $\tau^{\rm ex}$ is allowed to break conservation of total angular momentum. We note that $\tau^{\rm ex}$ can also include a friction term: $\tau^{\rm ex} = \tau -\Gamma^\Omega \Omega$. 
In 2D the only vector (of an assumed isotropic system) is $\vecv$, hence $\boldsymbol{C} = \eta^C(\vecr,t) \vecv$, where $\eta^C$ can be a function of $\vecr$ and $t$. Symmetry (Curie's principle~\cite{Mazur}) also tells us that $\eta^C = \eta^C(\rho,\ell,\vecv^2)$. 
Here, the antisymmetric part of the CM stress appears in the SAM dynamics as required from the balance of angular momentum~\cite{NJP,Mazur}. Note that in principle, active stresses can create antisymmetric stresses that are not compensated by the appropriate term in the $\ell$ dynamics. However, within our formalism, these terms do not appear in ${\bm \sigma}$, instead they appear in $\boldsymbol{f}$. Then using $ \varepsilon_{ij}\sigma^c_{ij} = \Gamma(\Omega-\omega^c) - 2\eta^A \nabla \cdot \vecv^c$, the dynamics for the SAM reads
\begin{eqnarray}
\label{eq:tam1}
\nonumber\dot{\ell} + \nabla_j (v^c_j \ell) &=& \tau^{\rm ex} - \Gamma(\Omega-\omega^c) + 2\eta^A \nabla \cdot \vecv^c \\
&+& \nabla \cdot (\eta^C \vecv^c) \, .
\end{eqnarray}

It is instructive to write the total momentum dynamics that results from Eqs.~\eqref{eq:cm_stress} and \eqref{eq:tam1}, which assumes the following form (see Appendix \ref{app:cm_to_total}):
\begin{eqnarray}
\label{eq:total_momentum_all}
\nonumber& &\sigma_{ij} = -P\delta_{ij}  +v_k\nabla_l \left[\chi_{ijkl} + \frac{\ell}{2} \varepsilon_{jl}\delta_{ik} + \frac{\eta^C}{2} \varepsilon_{il}\delta_{jk}\right] \\
& &+ \tilde{\eta}_{ijkl} \nabla_l v_k + \half\varepsilon_{ij}\tau^{\rm ex}   \, ,
\end{eqnarray}
with
\begin{eqnarray}
\label{eq:total_momentum_viscosity}
\tilde{\eta}_{ijkl} =  \eta^e_{ijkl} + \frac{\eta^o}{4}\gamma^o_{ijkl} + \eta^B \varepsilon_{lk}\delta_{ij} +\frac{\eta^C-\ell}{2} \varepsilon_{il}\delta_{jk} \, . \,\,
\end{eqnarray}
This is the most general total momentum stress tensor using only structural symmetry arguments.

Equations~\eqref{eq:general_cm_stress}, \eqref{eq:cm_stress} and \eqref{eq:tam1} seem to provide the most general CM dynamics of isotropic chiral fluids. In the following subsection we show that there are other constraints on the form of the total momentum stress tensor of Eq.~\eqref{eq:total_momentum_all} that should be taken into account. These constraints are shown to restrict the form of the `odd' terms.

\subsection{Symmetry is not the only constraint} \label{sec:symm}

So far we have only considered `structural' symmetry arguments. There are two other constraints that one may impose: (i) Galilean invariance and (ii) a symmetric stress tensor for the total momentum. 

It is not clear to what extent Galilean invariance can actually be imposed on active materials as there are examples in which it is broken (i.e., self advection of polar order parameter~\cite{PRX2021, NJP, Marchetti2013} and also interactions with surfaces~\cite{Maitra2020}). Nevertheless, assuming activity from interactions with surfaces is encompassed in $\boldsymbol{f}$ and $\tau^{\rm ex}$, and if we assume that all other interactions only depend on position and relative velocities, Galilean invariance can only be broken by $\boldsymbol{f}$ or $\tau^{\rm ex}$. 

The second requirement is completely general, and is nothing more than the statement that (disregarding $\boldsymbol{f}$) the total momentum is conserved (we note again that active forces that conserve total momentum are included in the stress while those that are not are written in $\boldsymbol{f}$).
With this in mind, the stress associated with the total momentum $g_i  = g^c_i + \frac{1}{2} \varepsilon_{ij}\nabla_j \ell$  must be expressible as a symmetric tensor, and the total momentum dynamics should be Galilean invariant (up to $\boldsymbol{f}$). The CM stress, however, does not have these two constraints. 

As shown in Appendix~\ref{app:total} we find that  for the total momentum stress to be symmetric we must have:
\begin{eqnarray}
\label{eq:eta_relation}
\eta^B = \bar{\eta}^B + \half\left(\eta^C -\ell\right)  \, ,
\end{eqnarray}
where $\bar\eta^B$ couples rotation (i.e., the fluid vorticity) to pressure, while $\eta^B$ also contains a part that is related to the odd viscosity that couples the two shear directions~\cite{VitelliKubo}.
In order to have Galilean invariance we must also have
\begin{eqnarray}
	\label{eq:chi}
	\chi_{ijkl} = - \frac{\eta^C}{4} \left(\gamma^o_{ijkl} - 2\varepsilon_{lk}\delta_{ij} \right)  \, .
\end{eqnarray}
%
%
Taken together, the total momentum stress is
\begin{eqnarray}
\label{eq:total_momentum_with_relation}
\nonumber	\sigma_{ij} &=&   -P\delta_{ij}  + \half\varepsilon_{ij}\tau^{\rm ex} \\
&+& \left[ \eta^e_{ijkl} + \frac{\tilde{\eta}^o}{4} \gamma^o_{ijkl} + \bar{\eta}^B \varepsilon_{lk}\delta_{ij}\right] \nabla_l v_k \, .
\end{eqnarray}
This result leads us to our first important conclusion: the odd part of the viscosity looks like the well-known odd viscosity~\cite{Avron1998,Banerjee2017,MarLub2021,MarLub2024} with a modified odd viscosity, $\tilde\eta^o = \eta^o + \eta^C-\ell$, and an `odd pressure' term $\bar\eta^B$~\cite{Abanov2021}. 
Because ${\bm \sigma}$ is related to the total momentum, this is the stress that will be measured in, e.g., a rheology experiment. Then, the forces exerted on a boundary of an odd fluid are of the form of Eq.~\eqref{eq:total_momentum_with_relation}. We remark again that this does not include any potential external surface forces that are included in $\boldsymbol{f}$. 


Taking into account the constraints of Eqs.~\eqref{eq:eta_relation} and \eqref{eq:chi}, the most general CM stress is: 
%
\begin{eqnarray}
	\label{eq:CM_stress_final}
	\nonumber& &\sigma_{ij}^c = -P\delta_{ij} + \varepsilon_{ij}\frac{\Gamma}{2} (\Omega-\omega^c) \\
	\nonumber& &-v_k^c \nabla_l \left[  \frac{\eta^C}{4} \left(\gamma^o_{ijkl} - 2\varepsilon_{lk}\delta_{ij} \right)   \right]\\
	\nonumber& & + \bigg[ \eta^e_{ijkl} + \frac{\eta^o}{4}\gamma^o_{ijkl} - \eta^A\varepsilon_{ij}\delta_{kl}  \\
	& &\qquad\qquad+ \left( \frac{\eta^C-\ell}{2} + \bar\eta^B  \right) \varepsilon_{lk}\delta_{ij} \bigg] \nabla_l v^c_k \, ,
\end{eqnarray}
which can also be written with the help of Eq.~\eqref{eq:tam1} as
\begin{eqnarray}
	\label{eq:CM_stress_final_ell}
	& &\sigma_{ij}^c = -P\delta_{ij} + \half\varepsilon_{ij} \left(\tau^{\rm ex} - \dot{\ell}\right) \\
	\nonumber & & -v_k^c \nabla_l \left[  \frac{\ell}{4} \left(\gamma^o_{ijkl} - 2\varepsilon_{lk}\delta_{ij} \right)  \right] \\
	\nonumber& & + \bigg[ \eta^e_{ijkl} + \frac{\tilde\eta^o}{4}\gamma^o_{ijkl}    +  \bar\eta^B   \varepsilon_{lk}\delta_{ij} \bigg] \nabla_l v^c_k \, ,
\end{eqnarray}
where we have used Eq.~\eqref{eq:etaB_switch}. The third term seems to break Galilean invariance, but it can be verified that when a boost (i.e., $\vecr \to \vecr - {\bf u}t$ where ${\bf u}$ is a constant) is also taken within the $\ell$ field, Galilean invariance is obeyed also in the CM.

It is useful to identify the type of interactions that lead to the various ``odd'' coefficients. By construction, the origin  of $\eta^o$, $\eta^A$, and $\eta^B$ is two-body interactions between particles CM positions (central-force interactions), while $\eta^C$ originates in spin-spin interactions (non-central force interactions). 
%
Examining the 
non-interacting results of Sec.~\ref{sec:pb}, we can separate the interacting and non-interacting contributions to the odd viscosity and specifically to $\eta^B$. In the non-interacting case we have $\eta^o=\eta^C=\eta^A=\eta^B=0$, in which case according to Eq.~\eqref{eq:eta_relation} we have $\bar\eta^B=\ell/2$ and Eq.~\eqref{eq:CM_stress_final} reduces to Eq.~\eqref{eq:pb_cm_stress}. In the general interacting case, $\bar\eta^B = \bar\eta^{\rm BI} + \bar\eta^{\rm BNI} = \ell/2 + \bar\eta^{\rm BNI}$ where  $\bar\eta^{\rm BI}$ and $\bar\eta^{\rm BNI} = \ell/2$ are the interacting and non-interacting parts of $\bar\eta^B$, respectively, such that 
$\eta^{B} = \eta^C/2 + \bar\eta^{\rm BI}$. There is no restriction on the value of $\bar\eta^{\rm BI}$.

Importantly, this means that the two-body central-force interactions are strongly coupled to the spin-spin interactions. It tells us that particles do not only exchange spins, but by doing so they also affect the vorticity of the fluid, i.e., the orbital angular momentum. This is of course not surprising, because even in a simple two-particle collision of rotating particles, some of the internal rotational energy is converted to orbital angular momentum~\cite{odd_ideal_gas}.

%
%

\section{CM stress after elimination of SAM}\label{sec:elim}

Because $\ell$ is not hydrodynamic~\cite{MarLub2021}, it seems reasonable that  $\ell$ can be eliminated and that the  CM dynamics, after integrating out $\Omega$, is sufficient to describe the system dynamics. In such case, it is also expected that the CM and total momentum dynamics coincide~\cite{Martin}. Although in normal systems this is indeed the case, we show in this section that in chiral active fluids, where $\ell$ relaxes to a non-vanishing value in the hydrodynamic limit, the CM description is generally insufficient as it does not obey the balance of total angular momentum, and its stress is not the same as the total momentum stress~\cite{MarLub2024}.

We eliminate the SAM  by solving Eq.~(\ref{eq:tam1}) in the hydrodynamic limit (long-wavelength low-frequency).  The solution that is detailed in Appendix~\ref{app:elimination} is valid for any ${\bm \tau}$, even if it is inhomogeneous and not constant in time. It essentially sets $\dot{\Omega}=0$ such that with the aid of the continuity equation we find that
\begin{eqnarray}
\label{eq:dot_ell}
\dot{\ell} \simeq -\tilde{I}\Omega_0 \nabla\cdot(\rho_0\vecv) = -\ell_0 \nabla\cdot\vecv - \tilde{I}\Omega_0 \vecv\cdot\nabla\rho_0  \, ,
\end{eqnarray}
where $\ell_0 = I_0 \Omega_{0}$ and $\Omega_0 = \tau(\vecr,t) / \Gamma^T$ with $\Gamma^T = \Gamma + \Gamma^\Omega$.  Substituting Eq.~(\ref{eq:dot_ell}) into Eq.~\eqref{eq:CM_stress_final_ell} gives the low-frequency long-wavelength dynamics of the CM momentum after relaxation of ${\Omega}$. The CM stress tensor is then
\begin{eqnarray}
\label{eq:CM_stress_after_elimination}
\nonumber\sigma_{ij}^c &=&  - P\delta_{ij} + \half \varepsilon_{ij} \left( \tau^{\rm ex} - \frac{I_o}{2} \vecv^c\cdot\nabla\Omega^0 \right) \\
&+& \!\!  \left(\eta^e_{ijkl}+ \frac{\eta^o+\eta^C}{4}\gamma^o_{ijkl} + \bar\eta^{\rm BI} \varepsilon_{lk}\delta_{ij} \right) \nabla_l v^c_k   \, , 
\end{eqnarray}
where $\tau^{\rm ex} = \tau -\Gamma^\Omega \Omega$ and we have used $\bar\eta^B = \ell/2 + \bar\eta^{\rm BI}$.
Importantly, even after the relaxation of  SAM, the CM stress tensor still depends on the SAM density and does not obey Galilean invariance (unless $\Omega_0$ is homogeneous). 

After relaxation of the angular momentum, the viscosity of the CM stress is very similar to that of the total  momentum, Eq.~\eqref{eq:total_momentum_with_relation}, where the difference is the kinetic (non-interacting) contribution of Sec.~\ref{sec:pb} that is proportional to the SAM density $\ell_0$ (note that after relaxation $\ell \simeq \ell_0$ in Eq.~\eqref{eq:total_momentum_with_relation}). The viscosity of ${\bm \sigma}^c$ is purely a result of interactions, both central and non-central (spin-spin).

%

A remarkable conclusion is that whenever the SAM acquires a non-vanishing average value, the CM momentum does not completely describe the system dynamics, unlike the situation in passive fluids~\cite{Martin,MarLub2024}. This is seen from two different aspects: (i) The CM stress tensor is not equivalent to the total momentum stress, where even the viscosities differ (compare Eq.~\eqref{eq:CM_stress_after_elimination} and Eq.~\eqref{eq:total_momentum_with_relation}), and (ii) the appearance of angular velocity in the dynamics of the CM linear momentum contribute an antisymmetric stress (that is not $\half\varepsilon_{ij}\tau^{\rm ex}$), which breaks the balance of total angular momentum. 
%

%

\section{Discussion} \label{sec:discussion}

A very useful method of formulating field theories is by using conservation laws and system symmetries~\cite{LubenskyBook}. This gives relatively easy access to the phenomenology of the system in question. Progress in identifying the relevant `odd' terms in the viscosity tensor was recently done in both 2D~\cite{Abanov2021,VitelliKubo} and 3D~\cite{Khain2022}. This work relies on structural symmetries and the conservation of the CM momentum and identifies three odd terms in 2D: Odd (Hall) viscosity, odd pressure, and odd torque~\cite{Abanov2021}. 

In this paper we utilize another constraint, that was not accounted for in previous work, which is the fact that the total momentum stress tensor must be symmetric. Importantly, this total momentum stress (see Eq.~\eqref{eq:total_momentum_with_relation}) is the actual force on the system boundaries, and as we showed, it is not equal to the CM stress, even after relaxation of SAM (see Eq.~\eqref{eq:CM_stress_after_elimination}). After such relaxation (which happens in non-hydrodynamic times) the CM stress acquires a very similar form to the total momentum stress, where only two odd terms appear: odd viscosity ($\tilde\eta^o$) and odd pressure ($\bar\eta^B$). 
(The excitation spectrum of systems that includes these two odd terms was analyzed Ref.~\cite{MarLub2024}.)

Yet, the CM and total momentum stresses are not equivalent (even after SAM relaxation); The odd viscosity and odd pressure in the total momentum have an extra kinetic contribution $\sim\ell_0$~\cite{MarLub2024} compared to the CM stress. Furthermore, when the active torques are inhomogeneous the CM stress has an antisymmetric stress (which does not appear in the symmetric total momentum stress) that do not obey Galilean invariance and seems to break conservation of total angular momentum.  Total angular momentum is of course conserved, but it requires accounting for both the CM linear momentum and the SAM, even after SAM relaxation. This suggests that the CM momentum is not sufficient to describe a system in which the SAM relaxes to a non-vanishing value. 

The odd torque ($\eta^A$), which is allowed by (structural) symmetry, cannot be found using rheological experiments, nor in experiments that can access the CM stress in times longer than the relaxation of SAM. 
Nevertheless, in Ref.~\cite{VitelliKubo} a value for $\eta^A$ was found from molecular dynamic simulations (even though it was small) when a uniform deformation was imposed with periodic boundary conditions. This could be a consequence of the periodic boundary conditions together with measuring the CM stress at time-scales smaller than the relaxation time of the angular momentum density.

The odd viscosity that appears in the total momentum is $\tilde\eta^o = \eta^o + \eta^C - \ell_0$, which has three contributions:  (i) $\eta^o$ that results from two-body interactions between particles CM positions, (ii) $\eta^C$ that is a consequence of short-range spin-spin interactions, and (iii) a kinetic non-interacting contribution that depends on $\ell_0$ and is always present for spinning particles. Similarly, the odd pressure term $\bar\eta^B = \ell_0/2 + \bar\eta^{\rm BI}$ has two contributions: A kinetic contribution $\sim\ell_0$ that is always present and a contribution that results from two-body center-force interactions $\bar\eta^{\rm BI}$. As discussed above, the CM stress (after relaxation of SAM) contains only the interacting parts of the odd viscosity and odd pressure.

From the CM point-of-view $\eta^B$ is an odd pressure and has no relation with the spin-spin interaction $\eta^C$. However, another important outcome from imposing the symmetry of the total momentum is that it imposes a relation between the two-body center-force interactions and the spin-spin interactions $\eta^B = \eta^{\rm BI} + \eta^C/2$.  A relation of such is to be expected because when spinning particles collide they exchange both linear and angular momenta, where some of the internal rotational energy is converted to orbital angular momentum~\cite{odd_ideal_gas}. It also shows that $\eta^B$ also includes terms that are related to the odd viscosity rather than the odd pressure, which is ultimately $\bar\eta^B$.

To conclude, when treating {\it chiral active fluids} the CM momentum is not sufficient to describe the system's dynamics (see Eq.~\eqref{eq:CM_stress_after_elimination}). The proper description arises when considering the total momentum that accounts for all of the momentum, including rotations around the CM of a complex molecule. The stress related to this total momentum, which is  experimentally accessible, is quite different from the CM stress (although the `odd' terms have the same form, their magnitudes differ) and it can always be written as symmetric, which imposes another symmetry constraint on the viscosity coefficients. 
We hope that this work will modify some of the concepts in investigations of chiral active materials and specifically odd viscosity that were so far focused on the dynamics of the CM momentum.


\begin{acknowledgments} 
\emph{Acknowledgments:} 
TM acknowledges funding from the Israel Science Foundation (Grant No. 1356/22) and Grant No.2022/369 from the United States-Israel Binational Science Foundation (BSF).
TCL acknowledges funding from the National Science Foundation Materials Research Science and Engineering Center (MRSEC) at University of Pennsylvania (Grant No. DMR-1720530). 
TM and TCL thank Vincenzo Vitelli, Alexander Abanov, and Anton Souslov for useful discussions that ignited this work.
\end{acknowledgments}

\appendix

\section{From CM stress to total momentum stress} \label{app:cm_to_total}

In this appendix we derive Eq.~\eqref{eq:total_momentum_all} from Eqs.~\eqref{eq:cm_stress} and \eqref{eq:tam1}. We start by substituting Eq.~\eqref{eq:tam1} into Eq.~\eqref{eq:cm_stress}
\begin{eqnarray}
	\label{eq:CM_to_total_momentum}
	\nonumber\dot{g}^c_i &+& \nabla_j(v^c_jg^c_j)  =   \half \varepsilon_{ik} \nabla_k \nabla_j\left[ \left(\eta^C-\ell\right) v^c_j\right] \\ 
	\nonumber&+&\nabla_j\Big[ -P\delta_{ij}  + \left( \eta^e_{ijkl} + \frac{\eta^o}{4}\gamma^o_{ijkl} - \eta^B \varepsilon_{kl}\delta_{ij} \right) \nabla_l v^c_k \\
	&+& v^c_k\nabla_l \chi_{ijkl} + \half\varepsilon_{ij}\left(\tau^{\rm ex} - \dot{\ell} \right)  \Big] + f_i\, .
\end{eqnarray}
Then, we convert  the advection term
\begin{eqnarray}
	\label{eq:total_advection}
	\nonumber\nabla_j(v_jg_j) &=& \nabla_j\left[ \left(v^c_j + \frac{1}{2\rho} \varepsilon_{jk} \nabla_k \ell\right) \left(g^c_i + \frac{1}{2}  \varepsilon_{ik} \nabla_i \ell\right)\right] \\
	&=&\nabla_j(v^c_jg^c_i) +\half\left( \varepsilon_{ik} v^c_j + \varepsilon_{jk} v^c_i \right) \nabla_k \ell  \, ,
\end{eqnarray}
and  find Eq.~\eqref{eq:total_momentum_viscosity}.

\section{Total momentum stress constraints} \label{app:total}

In this appendix we derive the constraints due to the total momentum stress, Galilean invariance, and symmetry. 
Our starting point is Eq.~\eqref{eq:total_momentum_all}. It is convenient to proceed by converting the $\eta^B$ term of Eq.~\eqref{eq:total_momentum_all} to a more familiar form using 
\begin{eqnarray}
	\label{eq:etaB_switch}
	\nonumber & & \nabla_j\left[X \varepsilon_{lk}\delta_{ij} \nabla_l v_k\right] = \nabla_j \left(  X \varepsilon_{jk} \delta_{il} \nabla_l v_k \right) \\
	& &\qquad\qquad + \left(\varepsilon_{jk}\delta_{il} -\varepsilon_{lk}\delta_{ij}  \right) \nabla_j  \left( v_k \nabla_l X\right) \, ,
\end{eqnarray}
where $X(\vecr,t)$ is an arbitrary function. This allows us to write the total momentum stress as: 
\begin{eqnarray}
	\label{eq:total_momentum_not_const}
	\nonumber& &\sigma_{ij} = -P\delta_{ij} + \half\varepsilon_{ij}\tau^{\rm ex} + \Big[ \eta^e_{ijkl} + \frac{\eta^o}{4}\gamma^o_{ijkl}  \\
	\nonumber& &+ \bar{\eta}^B\varepsilon_{lk}\delta_{ij} + \left(\eta^B - \bar{\eta}^B\right) \varepsilon_{jk}\delta_{il} + \frac{\eta^C-\ell}{2} \varepsilon_{il}\delta_{jk} \Big] \nabla_l v_k \\
	\nonumber& &+ v_k\nabla_l \Big[\chi_{ijkl} + \frac{\ell}{2} \varepsilon_{jl}\delta_{ik} + \frac{\eta^C}{2} \varepsilon_{il}\delta_{jk} \\
	& &\qquad\qquad\qquad+\, \left(\eta^B - \bar{\eta}^B\right)  \left(\varepsilon_{jk}\delta_{il} -\varepsilon_{lk}\delta_{ij}  \right)  \Big]  \, .
\end{eqnarray}
Here we have introduced artificially $\bar{\eta}^B$, which is useful as discussed in the main text.
Clearly, to have a symmetric stress tensor we must have $\eta^B - \bar{\eta}^B= \half\left(\eta^C -\ell\right)$ such that the total momentum stress can be written as
\begin{eqnarray}
	\label{eq:total_momentum_stress_final}
	\nonumber\sigma_{ij} &=&  \left[ \eta^e_{ijkl} + \frac{\eta^o+\eta^C-\ell}{4} \gamma^o_{ijkl} + \bar{\eta}^B \varepsilon_{lk}\delta_{ij}\right] \nabla_l v_k \\
	\nonumber&-&P\delta_{ij}  + \, v_k\nabla_l \bigg[\chi_{ijkl} + \frac{\eta^C}{4} \left(\gamma^o_{ijkl} - 2\varepsilon_{lk}\delta_{ij} \right) \bigg] \\
	&+& \half\varepsilon_{ij}\tau^{\rm ex}  \, .
\end{eqnarray}
In order to have Galilean invariance we must also have $	\chi_{ijkl} = - \frac{\eta^C}{4} \left(\gamma^o_{ijkl} - 2\varepsilon_{lk}\delta_{ij} \right)$, which leads to Eq.~\eqref{eq:total_momentum_with_relation}.

\section{Elimination of SAM} \label{app:elimination}

Here we solve for $\Omega$ in the hydrodynamic limit. We start by writing ${\Omega} = {\Omega}_0 + \delta{\Omega}$ and $\rho = \rho_0 + \delta\rho$ while assuming that $\delta{\bm \Omega}$ and $\delta\rho$ are of the order of ${\bm\nabla}{\bf v}^c$. Here we include also possible friction with the surface such that $\tau^{\rm ex} = \tau -\Gamma^\Omega \Omega$. We then have:
\begin{eqnarray}
	\label{eq:Omega_0}
	& &\left(\partial_t + \Gamma^T / I_0 \right) \Omega_0 = \tau(\vecr,t)/I_0 \, , \\
	\label{eq:delta_Omega}
	& &\left(\partial_t + \Gamma^T / I_0 \right) \delta\Omega = \Phi(\vecr,t)/I_0 \, ,
\end{eqnarray}
where 
\begin{eqnarray}
	\label{eq:Phi}
	\nonumber \Phi(\vecr,t) &\equiv& \Gamma\omega^c - I_0\vecv^c\cdot\nabla\Omega_0 \\
	&+& 2\eta^A\nabla\cdot\vecv^c +\nabla\cdot\left(\eta^C\vecv^c\right) \, ,
\end{eqnarray}
$I_0 =  \tilde{I}\rho_0$, and $\Gamma^T = \Gamma + \Gamma^\Omega$. The solution to Eqs.~\eqref{eq:Omega_0}-\eqref{eq:delta_Omega} is
\begin{eqnarray}
	\label{eq:Omega_0_sol}
	& &\Omega_0(\vecr,t) = \Omega_0(t=0) \e^{-T} \!\! + \!\! \int_{0}^T \frac{\tau(\vecr,u)}{\Gamma^T} \e^{-(T-u)} \D u  , \quad\\
	\label{eq:delta_Omega_sol}
	& &\delta\Omega(\vecr,t) = \delta\Omega(t=0) \e^{-T} \!\!+ \!\!\int_{0}^T \!\! \frac{\Phi(\vecr,u) }{\Gamma^T} \e^{-(T-u)} \D u   , \quad
\end{eqnarray}
with $T\equiv \Gamma^T t / I_0$. In the hydrodynamic limit $T\gg 1$ and we are left with 
\begin{eqnarray}
	\label{eq:Omega_sol_final}
	\Omega_0(\vecr,t) \simeq \tau(\vecr,t) / \Gamma^T  \,\,\, ; \,\,\, \delta\Omega(\vecr,t) \simeq \Phi(\vecr,t) / \Gamma^T   \, .
\end{eqnarray}
This solution is valid for any ${\bm \tau}$, even if it is inhomogeneous and not constant in time.

\bibliography{odd_viscosity_2d_pre_citation.bib}

\end{document}